\documentclass[10pt,conference]{IEEEtran}
\IEEEoverridecommandlockouts
\pdfoutput=1 
\usepackage{cite}
\usepackage{amsmath,amssymb,amsfonts}
\usepackage{algorithmic}
\usepackage{graphicx}
\usepackage{textcomp}
\usepackage{xcolor}
\usepackage{url}

{
  \textbf{Example}\begin{itshape}%
}%
{\end{itshape}}

\newtheorem{definition}{Definition}
\newtheorem{prop}{Proposition}

\newcommand{\entropy}{{\mathcal H}}
\newcommand{\leak}{{\mathcal L}}
\newcommand{\mutual}{{\mathcal I}}

\makeatletter
\newcommand{\linebreakand}{%
  \end{@IEEEauthorhalign}
  \hfill\mbox{}\par
  \mbox{}\hfill\begin{@IEEEauthorhalign}
}
\makeatother

\def\BibTeX{{\rm B\kern-.05em{\sc i\kern-.025em b}\kern-.08em
    T\kern-.1667em\lower.7ex\hbox{E}\kern-.125emX}}
\begin{document}

\title{HyperGI: Automated Detection and Repair of Information Flow Leakage}

\author{\IEEEauthorblockN{Ibrahim Mesecan}
  \IEEEauthorblockA{\textit{Iowa State University} \\
      Ames, Iowa, USA \\
      imesecan@iastate.edu}
  \and
  \IEEEauthorblockN{Daniel Blackwell}
  \IEEEauthorblockA{\textit{University College London}\\
     London, UK \\
     daniel.blackwell.14@ucl.ac.uk}\\
  \and
  \IEEEauthorblockN{David Clark}
  \IEEEauthorblockA{\textit{University College London}\\
      London, UK \\
      david.clark@ucl.ac.uk}\\
  \linebreakand 
    \IEEEauthorblockN{Myra B. Cohen}
  \IEEEauthorblockA{\textit{Iowa State University}\\
      Ames, Iowa, USA \\
      mcohen@iastate.edu}
  \and
 \IEEEauthorblockN{Justyna Petke}
  \IEEEauthorblockA{\textit{University College London}\\
     London, UK \\
     j.petke@ucl.ac.uk}
    }
\maketitle
\begin{abstract}
  Maintaining confidential information control in software is a
persistent security problem where failure means secrets can be
revealed via program behaviors. Information flow control techniques
traditionally have been based on static or symbolic analyses --- limited
in scalability and specialized to particular languages. When programs
do leak secrets there are no approaches to automatically repair them
unless the leak causes a functional test to fail.  We 
present our vision for HyperGI, a genetic improvement framework that
detects, localizes and repairs information leakage. Key elements of
HyperGI include (1) the use of two orthogonal test suites,
(2) a dynamic leak detection
approach which estimates and localizes potential leaks, and (3) a
repair component that produces a candidate patch using genetic
improvement. We demonstrate the successful use of HyperGI on several programs which have no failing functional tests.
We manually examine  the resulting patches and identify trade-offs and future directions for fully realizing our vision.
\end{abstract}

\begin{IEEEkeywords}
information flow leakage, genetic improvement
\end{IEEEkeywords}
\section{Introduction}

The problem of software accidentally leaking confidential information is longstanding~\cite{Denning82}, much researched~\cite{SabelfeldM03}, and remains an ongoing problem~\cite{CherubinCP2019}. Its ubiquity and problematic nature has led to high profile security failures such as the famous Heartbleed Bug~\cite{Heartbleed2014}. The verification research community has extensively studied ensuring Information Flow Control (IFC) as part of the programming process over many decades~\cite{VolpanoIS96,Vassena_2021}. IFC is the problem of guaranteeing that a software and a security policy pair satisfy a security property.  

As security properties are safety properties most research into IFC has been via verification tools and static or symbolic analyses\cite{HeusserM10,Phan:2012}.  While dynamic approaches are not unknown~\cite{HedinSPS17, MagaRS2012} they have been comparatively neglected until recent years.  Contemporary software is often large and getting larger~\cite{GTB} and the recent rapid development in the ability of fuzzers to detect security related problems in software is causing a rethink about the value of dynamic approaches and their big advantages in scalability and flexibility~\cite{AFL}.  IFC has lacked significant uptake in industry, a significant exception being the SEL4 microkernel~\cite{SEL4}. Rather, the emphasis has been on discovering and patching exploitable security vulnerabilities. However, detecting information leaks can not only detect errors in code's flow logic but also functional errors that lead to leaks, such as memory leaks and buffer overflows~\cite{MalacariaTD16}.  In recent work, Mechtaev et al. demonstrated that they could automatically repair  the Heartbleed Bug\cite{mechtaev2016angelix}. However, we caution that this is a  special case of IFC, where the program can be made to crash when the safety property is violated. We cannot expect this to hold in general as we demonstrate later.

In this paper we take a fresh look at IFC and ask if we can use a dynamic approach to \textit{both detect and repair} this important type of security bug.  We make two assumptions. First, we cannot assume an IFC will cause the program to fail. Second, we realize that there could be a trade-off between maintaining the original program semantics and removing the information flow leakage.   We propose an end-to-end framework called HyperGI.  HyperGI, takes a program and a security policy and tests (technically, hypertests) the program for evidence that it leaks and, if it does, estimates the size of the leak. Then HyperGI uses Genetic Improvement~\cite{PetkeHHLWW17} to automatically repair the leak while attempting to minimise changes to the program semantics.  While the concept of hypertesting programs has been around at least since Kinder's work~\cite{Kinder2015}, it has been little explored in the software engineering community. One recent effort is CT-Fuzz where the hypertest oracle is observing timing and control flow path differences~\cite{HeEC2020}.  
The strong novelty in our approach is the use of quantified information flow estimates in leak repair, combined with more traditional test cases to ensure functionality invariance.

We have implemented a prototype of HyperGI and apply this to three programs (two of which reported security vulnerabilities used in prior research). We demonstrate that we can reduce the leakage while retaining most of the program functionality. However, we not only identify the need for a multi-objective approach, we note several key aspects of future work needed to fully realize HyperGI, such as the identification of quality test suites for IFC repair.

The contributions of this work are:
\begin{itemize}
\item A framework for dynamically detecting, quantifying and repairing information leakage;
\item A prototype implementation and first case study to demonstrate its potential.
\end{itemize}

 In the next section we present HyperGI along with background.
 We then present our study in sections \ref{sec:study} and \ref{sec:results}.

\section{HyperGI}

\begin{figure}[h]
    \centering 
    \includegraphics[width=3.2in]{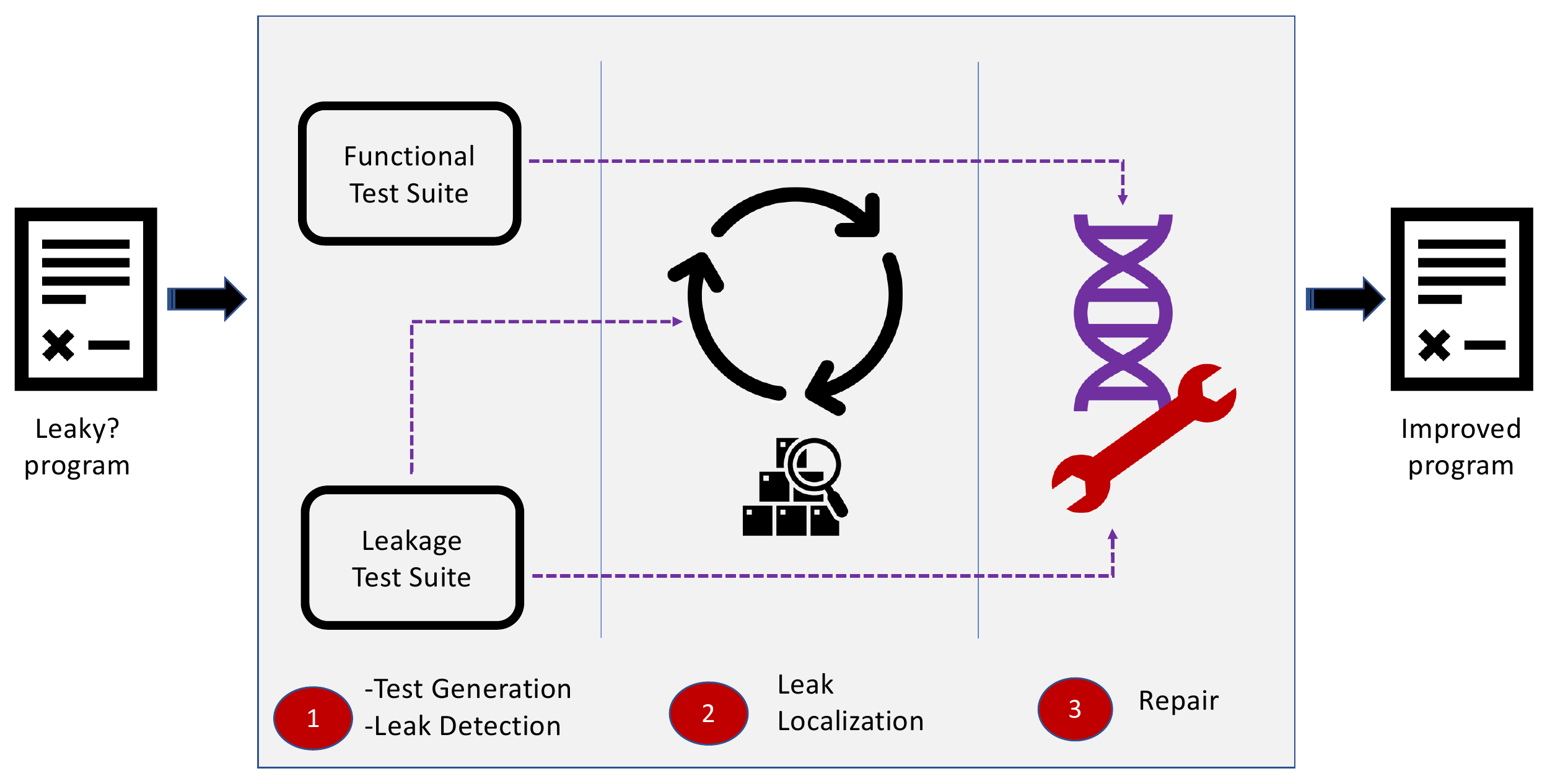}
    \caption{Overview of HyperGI which consists of three stages}\label{fig:overview}
\end{figure}

Figure \ref{fig:overview} shows a high-level overview of HyperGI. 
We start with a program (possibly) containing a leak and first generate two types of test sets, Hypertests and Functional tests.  
We then use a dynamic analysis with just the Hypertest test suite to localize the area of leakage in the program.
We then use genetic improvement, using both test suites to iteratively improve the program. 
We describe each step in more detail next, but first we provide  an overview of noninterference and hyperproperties which are fundamental to realizing HyperGI.

\subsection{Preliminaries}
\label{sec:preliminaries}

A property of program executions can be associated with a partition of the set of all executions into those that have the property and those that do not. 
Some properties cannot be expressed using an individual execution; rather they require two or more. 
An example is the \textit{noninterference} property which states  
  \textit{in any software system in which users are divided into groups with different information security access privileges, low security users should not be aware of the actions of  high security users} \cite{GoguenM82a}. 
We focus on input/output noninterference for imperative, deterministic programs \cite{SabelfeldM03}.    There are two groups of users, high security and low security. The states of the program 
are partitioned between high and low users and low users cannot control inputs to or read from high users' variables. 
A program contains the \textit{input/output noninterference property}  iff, for every pair of initial states with the same values of all low variables, but different values for high variables, the post execution states also have the same observable values, i.e., no information about the high variables is \textit{leaked}.  
In practice, leaks occur when data from memory is revealed as in a buffer overflow and/or when there is data dependency in control flow. 

We have created an exemplar Triangle program to demonstrate.  
It accepts 3 integers representing the length of each side of a triangle. 
The first side is a secret value. 
It returns the type of triangle (isosceles, scalene or equilateral).

{\footnotesize
\begin{verbatim}
TriangleType typeOf(int high, int low1, int low2){
  if (h == low1 && low1 == low2)
    return EQUILATERAL;
  else if (high==low1 || high==low2 || low1==low2)
    return ISOSCELES;
  else
    return SCALENE; }
\end{verbatim}
}
Due to the \verb|if| statements comparing the secret side, any return value potentially reveals some information about the secret.  The input \verb|{(high=?, low1=3, low2=4)}| returning \verb|SCALENE| indicates that the secret value \verb|high| was neither 3 nor 4, but is only observable if a second test using a different secret returns a different value.    If the input \verb|{(high=?, low1=3, low2=4)}| returned \verb|ISOSCELES|, then the user can deduce that the secret is either 3 or 4. 
Observing different outputs for the same low inputs but different secret is a violation of the noninterference property.
To detect this type we need to use \textit{hypertests}. 
Each hypertest is a set of inputs that satisfy the initial states specification of noninterference: the low part of the initial states are the same and the high parts are different. 
We use the notion of Quantified Information Flow (or QIF) to measure leakage.

\vspace{-0.15cm}
\subsection{Stage 1: Test Generation and Leak Detection}
\label{TestGenerationandLeakDetection}
In the Triangle program there are two (possibly competing) notions of correctness, functional and noninterference.
This program is functionally correct (assuming it accepts only valid triangle inputs); program repair techniques cannot help since there are no failing tests. A set of hypertests  may expose the leakage, but fixing the leak can impact functional correctness.  A key feature of HyperGI is the use of two independent test suites, one used to test correct program semantics and a hypertest suite used to measure information leakage.

Generating hypertests is challenging since the input space for finding inputs that expose noninterference may be enormous. 
HyperGI uses a binary search-like algorithm to solve this problem. 
It simultaneously \textit{detects} noninterference violations and \textit{generates} tests to maximize the information leakage. 
It starts by halving the input space and selects a number (parameter) of low inputs from each half. 
It  then runs the program with a number of executions (another parameter)  which alters only the high input(s). Last it checks the resulting output (or data in memory, depending on the security policy) and measures if the same/different values are returned.    It builds a priority queue to store hypertests that detect leakage (prioritized by high to low leakage). 
At each iteration, HyperGI chooses the half of the input space with the largest overall leakage (across all inputs) and repeats on that half.  When complete, if it detected a leak, it also has a set of hypertests that reveal it.


 

 

\subsubsection{Quantified Information Flow (QIF)}
Based on Denning \cite{Denning82}, Clark et al. constructed the first program analysis using an information theoretic framework to measure the size of leaks \cite{ClarkHM01}; when a leak is of size zero, noninterference holds. 
Both the program language security  and machine learning communities have extensively researched estimating information quantities \cite{AlvimCMMPS2020,Szabo2014}. Bounding QIF can be expressed as a hyperproperty \cite{HeusserM10,ClarksonS10} and has been shown to be PSPACE-hard to verify for exact values \cite{YasuokaT14}. 
HyperGI uses comparative estimations of entropy aiming to reduce the size of the leak to $0$. We briefly sketch the  mathematical framework .

Given a random variable, the entropy of the random variable is a statistic of its underlying probability distribution that captures the quantity of disorder in the distribution.

\begin{definition}
Let $X$ be a random variable, let $x$ range over the events of $X$, and let $p(x)$ be the probability distribution of $X$. The entropy of $X$,  $\entropy(X)$, is defined as follows:

\vspace{-0.1cm}
\[ \entropy(X) = - \sum_{x \in X} p(x)\; log\; p(x)\]
\vspace{0.1cm}
where logs are usually base 2 to retrieve a value in bits.
\end{definition}

From Clark et al. we give a definition of the leak size \cite{ClarkHM01}.

\begin{definition}
Let $\langle H,L \rangle$ be the joint random variable in the initial states of a program with a security policy $L \sqsubseteq H$, where $H$ represents high security variable values and $L$ the low security variable values. Similarly, $\langle H',L' \rangle$ represents the joint random variable in the final states. The quantity of leakage from $H$ to $L'$, $\leak(L')$,  is:

\vspace{-0.4cm}
\[ \leak(L') = \entropy(L' | L) \] 
\end{definition}

\vspace{-0.2cm}
The intuition here is that the only source of information in the program executions is the input state, represented by the random variable $\langle H,L \rangle$. 
The entropy in $L'$ can only exist as a result of the program executions on the initial states, so after factoring out entropy due to $L$, any remaining entropy in $L'$ must be due to $H$. 
It can be shown that, for non-deterministic programs, this is equivalent to the mutual information between $H$ and $L'$ given knowledge of $L$, $\mutual(H ; L' | L)$ \cite{ClarkHM01}. 

Conditional entropy calculation is cumbersome, the following chain rule can streamline it for joint random variables \cite{Cover:2006}.

\vspace{0.1cm}
\begin{prop} Chain Rule for Entropy
\vspace{-0.2cm}
\[ \entropy(A,B) = \entropy(A | B) + \entropy(B) \]
\end{prop}

\vspace{-0.1cm}
\noindent We then have QIF.   $\entropy(L' | L) = \entropy(L', L) - \entropy(L)$.

\subsection{Stage 2: Leak Localization}\label{subsec:localization}
HyperGI uses a dynamic algorithm that iteratively removes each line of the program and calculates the change in QIF of the program with that line removed. 
Non-compilable programs have a change of zero.  
It normalizes all of the QIFs (dividing all by the maximum change) and partitions the resulting statements into equivalence classes.  
Probabilities are assigned to each class which guides the repair towards those statements which are likely to reduce information flow the most. 

\subsection{Stage 3: Repair}\label{subsec:repair}
\label{sec:repair}
We implemented genetic programming (GP)~\cite{Koza:1993} search on top of an existing genetic improvement framework, PyGGI \cite{An:2019}.  Chromosomes are patches to the AST. 
We use the standard, delete, replace, insert operators, as well as two new operators.  
Since information flow leakage is highly control flow dependent we added operators to insert new control flow. 
One creates new \verb |if| statements (using variables from the program) and the other creates a new \verb |for| loop.  
Statements within the if/for blocks are created by copying existing statements from the target program, or by creating simple assignments between existing program variables. 
Fitness is one of the essential parts of GI and it guides the search process by measuring how fit the patches are.
The HyperGI fitness function combines both the QIF and functional correctness.
The  \textit{fail rate} of mutant $k$ is \\ 
$fr_k = (\# failing \ functional \ tests) / (\# functional \ tests)$.
And, $l_o$ is the initial program leakage and $l_k/l_o$ is the normalized leakage of mutant $k$ (defined if mutant $k$ compiles and runs).
Then, the fitness of mutant $k$ can be defined as: \\ 
$f_k = 0.5 * l_k/l_o + 0.5 * fr_k$.

\section{Evaluation}
\label{sec:study}

We conducted a feasibility study to understand the potential for HyperGI. 
We answer the following research questions:

\noindent\textbf{RQ1} \textit{How does HyperGI compare with fuzzing in terms of leak detection?}

\noindent\textbf{RQ2} \textit{How well does HyperGI remove information flow leakage while maintaining software functionality?}

To answer these questions we conducted a pilot study: we first run fuzzers to gather functional tests; next, we run our binary search to generate hypertests; then, we run GP-based repair to try to decrease leakage; and, finally,  we manually analyse generated patches.

We use three C subjects, two of which were used in prior work on statically finding information flow leakage~\cite{HeusserM10} and which are simplified versions of the original programs from the CVE vulnerability database~\cite{CVE2009AppleTalk, CVE2007Underflow}. 
The third program is one that we wrote to demonstrate the second type of leakage described in this paper, a control-flow based privacy leak. 
We generate functional and leakage test suites, as described below. 
  We show details of the subjects in Table~\ref{tab:subjects}, and
  present security policies for the two new subjects:

\begin{table}
  \caption{Study subjects. For each we give the reference, the CVE number, the number of functional tests and the number of hypertests in our test suites. }
\label{tab:subjects}
\setlength{\tabcolsep}{5pt}
  \begin{tabular}{|l|l|c|c|c|}
    \hline
    \bf{Subject}& \bf{Ref} & \bf{CVE-\#}&  \bf{\# Funct} & \bf{\# Hyper}\\
    &&&\bf{Tests}&\bf{Tests}\\
    \hline
 Triangle (triangle) & -- & --& 234&194
    \\
    \hline
Apple Talk (atalk) &~\cite{HeusserM10}&CVE-2009-3002&297&255	\\
    \hline
Underflow (underflow) &~\cite{HeusserM10}&CVE-2007-2875& 186&100	\\
    \hline
\end{tabular}
\end{table}

\begin{tiny}
\begin{verbatim}
atalk:
  static int atalk_getname(struct socket *sock, struct sockaddr *uaddr,  int peer);
  Low Input: sock and peer
  Low Output: uaddr and the return from the function
  High (secret): Information in memory not available to the user

underflow:
  int underflow(int h, ll ppos);
  Low Input: ppos
  Low Output: function result;
  High (secret) function output: h  // original program leaks machine information
\end{verbatim}
\end{tiny}

For comparison with existing dynamic approaches, we ran two state-of-the-art fuzzers  AFL~\cite{AFL} and LibFuzzer~\cite{LibFuzzer} on all subjects to see if we could detect the leakage. 
Information leaks due to buffer overflows can often be found with fuzzers  as memory leaks or buffer overflows (such as Heartbleed) can be interpreted as crashes via tools such as  AddressSanitizer~\cite{AddressSanitizer}. 
We ran 5 runs of each (with different random starting seeds) for 24 hours for each program. 
To increase test input diversity, we also ran each fuzzer 20 times for 2 hours with randomly generated input seeds.
For functional testing we used all tests from all 50 runs (= 25 x 2
fuzzers) for each program with duplicates removed (see Table
\ref{tab:subjects} for final counts). 
To generate hypertests, we ran our binary search on each subject, as described in Section~\ref{TestGenerationandLeakDetection}.

To fix detected leaks, we ran GP for 25 epochs, each with 50 generations and a population of 32. 
The target fitness was 0.0 (the program ends if it reaches this fitness) and we examined the best solution (or the stopping solution in case the program ended before 50 generations). 
We examined all 75 patches (= 25 x 3 programs) for their quality and
categorized them based on how well they fix the leak and/or retain the
functional correctness of the program.  
For \texttt{atalk} and \texttt{underflow} we have developer patches from github for reference.

\section{Results}\label{sec:results}

Both fuzzers ran to time limits without finding any crashes or leakage. 
In contrast, our binary search approach generated hypertests
which were able to detect information flow leakage in all three
programs.
This indicates that these types of leakage cannot always be found by
conventional fuzzing. After 320 hours they were unsuccessful while the
binary search detected each leak in less than two hours.
In answer to \textbf{RQ1}: \textit{Our proposed binary search strategy was able to detect leaks for all three programs, in contrast to traditional fuzzing.}

Table~\ref{tab:results} shows the results of applying HyperGI to
reduce information flow leakage.
The first column shows the initial detected information flow leakage (i.e., $l_o$). 
As we can see, all three programs have a leak with QIF ranging from 13.00  in \texttt{atalk} to 0.83 bits in the \texttt{triangle} program.  
For each subject we show the median, average and standard deviation of
the raw (not normalized) QIF (i.e., $l_k$), the test failure ratio, (i.e., $fr_k$), and the overall program fitness (where 0 is the optimal fitness). 
For \texttt{atalk} we were able to reduce the QIF to zero while maintaining program functionality (0 failed tests). 
For the other two subjects we see a trade-off. 
In fact, in the \texttt{triangle} case we can never
create a semantically equivalent and non-leaking program.

\begin{table}

  \caption{The starting QIF, final QIF (not normalized), the test failure
    ratio and total fitness (with normalized QIF)  by subject
    (TR=triangle, AT=atalk,UF=underflow). Median (med), Average (avg) and Standard Deviation (std)}

  \label{tab:results}
  \centering
  \footnotesize
  \begin{tabular}{|l|c||c|c|c||c|c|c||c|c|}
    \hline
  &\textbf{Init.} &\multicolumn{3}{|c||}{\textbf{Post Patch}}	&\multicolumn{3}{|c||}{\textbf{Functional Fail}} &\multicolumn{2}{|c|}{}\\	
  &\textbf{QIF} &	\multicolumn{3}{|c||} {\textbf{QIF}}         &\multicolumn{3}{|c||} {\textbf{Ratio}}   &\multicolumn{2}{|c|}{\textbf{Fitness}}\\
    \hline
    & & Med&	Avg & 	Std &   Med& Avg & 	Std & 	Avg &Std\\
  TR &0.8&0.0&0.3&	0.5      &0.3	&0.3&	0.2       &0.3&	0.4 \\
    \hline
AT&13.0&0.0	&0.0 &	0.0	&0.0&0.0 &	0.0 &	0.0&	0.0      \\
    \hline
UF &5.4& 0.0&	2.1&	2.7 & 1.0 &	0.6&	0.5 &	0.5 &	0.0 \\
                                                                      
    \hline
\end{tabular}
\vspace{-0.2cm}
\end{table}

\begin{table}
  \caption{Results of manual inspection of generated patches, with respect to developer fixes. Shows percentage of patches.}
  \label{tab:manual}
\setlength{\tabcolsep}{3pt}
  \centering\footnotesize
  \begin{tabular}{|l|r|r|r|}
\hline
\textbf{Patch Quality} & \textbf{triangle} & \textbf{atalk} & \textbf{underflow}\\
\hline
    Semantically-equivalent fix &--&76\%&28\%\\
    \hline
    Leakage reduction, no functionality loss &0\%&24\%&0\%\\
    \hline
    No leakage, but functionality loss &64\%&0\%&28\%\\
    \hline
    Leakage reduction, loss of functionality &24\%&0\%& 4\%\\
    \hline
    No improvement over the original program & 8\%&0\%& 24\%\\
    \hline
    Introduced indeterministic behavior &4\%&0\%&16\%\\
    \hline

  \end{tabular}
  \vspace{-0.2cm}
  \end{table}

\begin{figure}
\centering
    \includegraphics[width=2.68in]{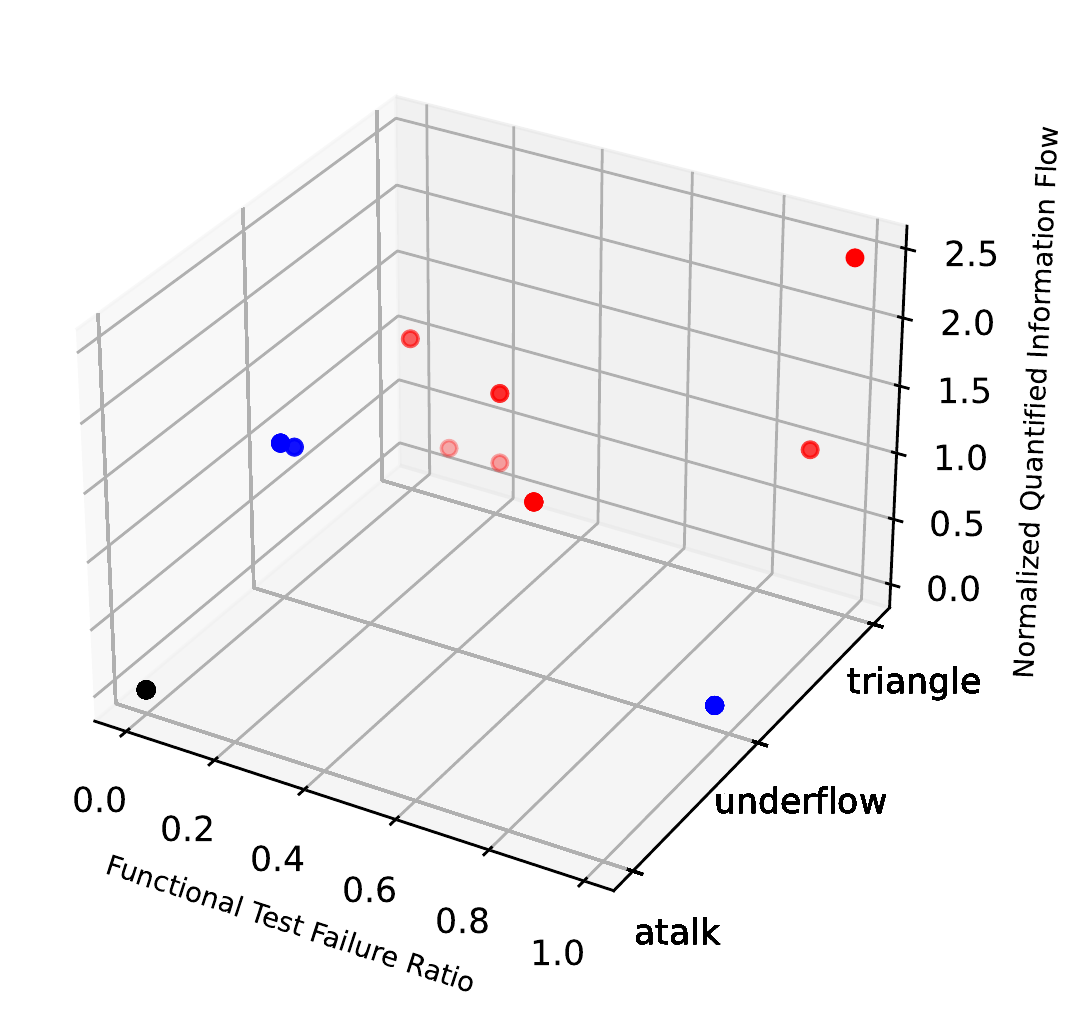}
    \caption{Scatter plot of Normalized QIF vs. test failure ratio. Lighter dots contain fewer points. QIF of 1 corresponds to leakage of the original program.}\label{fig:scatter}
\vspace{-0.4cm}
\end{figure}

We now turn to Figure~\ref{fig:scatter}.
~This shows the normalized QIF (over the pre-patch QIF)
versus the functional test failure ratio. 
We plot all 25 epochs for each program.  
For \texttt{atalk} (black), all 25 data have the same value thus showing a single point (all tests pass and QIF=0). 
For \texttt{underflow} (blue) we have two general patterns: either the
leakage is 0 and 97\% of functional tests are failing; or the leakage is 1 (no improvement), but the functionality is retained. 
This suggests the need for multi-objective optimization.  
For the \texttt{triangle} program (red dots) we see a wider range of points. 
Noticeably, there are no points which have retained all of the program
functionality.

We manually verified the quality of generated patches. 
For \texttt{triangle} we don't have a developer patch, but for the other two we used that as a baseline. 
Table~\ref{tab:manual} shows this data.
For two programs we find patches that are semantically-equivalent to the developer one (for 76\% of epochs for \texttt{atalk} and 28\% of epochs for \texttt{underflow}).
We also reduce leaks in the remaining 24\% of \texttt{atalk} runs, without loss of functionality.
For 84\% of cases for \texttt{triangle} and 32\% of cases for \texttt{underflow} we improve leakage at the cost of functionality.

A patch that reduces leakage, but breaks some program functionality can still be acceptable. The semantically-equivalent patches for \texttt{underflow} indeed fail our functional tests --- that is because to remove leakage the developers had to amend functionality of the program. 
This trade-off was not neccessary for \texttt{atalk}.

We found that some patches introduced nondeterministic behavior. 
In those cases our tests became flaky, and the QIF could potentially increase, as in the case of \texttt{triangle} in Figure~\ref{fig:scatter}.
Another example, from the \texttt{underflow} experiment, is a patch that removes a return statement, hence the program's
output became undefined and thus returned different results each time
it was run. 
In answer to \textbf{RQ2}: 
\textit{HyperGI was able to find patches semantically-equivalent to developer fixes. It found patches reducing leakage in all three programs.}

\section{Conclusions and Future Work}
\label{sec:conclusions}
We propose HyperGI, a framework for dynamically detecting, quantifying and fixing information flow leaks using lightweight dynamic analysis, hypertesting, and genetic improvement.
HyperGI was able to reduce information leakage in three programs, producing fixes semantically-equivalent to developer patches. We see a trade-off between quantified information flow and program functionality. 
As future work we will build HyperGI as a multi-objective framework. 
We found in some cases, e.g. \texttt{atalk,} despite obtaining a zero (or optimal) fitness, some patches were overfitted. Finding a good set of test cases, hypertests in particular, is important future work. Finally, we will run larger scale experiments on more subjects. 

\bibliographystyle{IEEEtran}
\bibliography{ms}
\end{document}